\documentclass[twocolumn, secnumarabic, amssymb, superscriptaddress, nobibnotes, aps, prd]{revtex4-2}

\usepackage{graphicx}
\graphicspath{{Figures/}} 

\usepackage{dcolumn}
\usepackage{bm}

\usepackage[utf8]{inputenc}
\usepackage[T1]{fontenc}
\usepackage{mathptmx}

\usepackage{upgreek} 
\usepackage{xcolor} 

\newcommand{\Mark}[1]{#1}

\begin{document}

\title[Mitigation of B-field Susceptibility TESs]{Mitigation of the Magnetic Field Susceptibility of Transition Edge Sensors \\ using a Superconducting Groundplane}

\author{M. de Wit}
 \affiliation{NWO-I/SRON Netherlands Institute for Space Research, Niels Bohrweg 4, 2333 CA Leiden, The Netherlands} 
 \email{M.de.Wit@sron.nl}
\author{L. Gottardi}
 \affiliation{NWO-I/SRON Netherlands Institute for Space Research, Niels Bohrweg 4, 2333 CA Leiden, The Netherlands} 
\author{M.L. Ridder}
 \affiliation{NWO-I/SRON Netherlands Institute for Space Research, Niels Bohrweg 4, 2333 CA Leiden, The Netherlands} 
\author{K. Nagayoshi}
 \affiliation{NWO-I/SRON Netherlands Institute for Space Research, Niels Bohrweg 4, 2333 CA Leiden, The Netherlands} 
\author{E. Taralli}
 \affiliation{NWO-I/SRON Netherlands Institute for Space Research, Niels Bohrweg 4, 2333 CA Leiden, The Netherlands} 
\author{H. Akamatsu}
 \affiliation{NWO-I/SRON Netherlands Institute for Space Research, Niels Bohrweg 4, 2333 CA Leiden, The Netherlands} 
\author{D. Vaccaro}
 \affiliation{NWO-I/SRON Netherlands Institute for Space Research, Niels Bohrweg 4, 2333 CA Leiden, The Netherlands} 
\author{J.-W.A. den Herder}
 \affiliation{NWO-I/SRON Netherlands Institute for Space Research, Niels Bohrweg 4, 2333 CA Leiden, The Netherlands}
 \affiliation{Anton Pannekoek Institute, University of Amsterdam, Science Park 904, 1098 XH Amsterdam, the Netherlands} 
\author{M.P. Bruijn}
 \affiliation{NWO-I/SRON Netherlands Institute for Space Research, Niels Bohrweg 4, 2333 CA Leiden, The Netherlands} 
\author{J.R. Gao} 
 \affiliation{NWO-I/SRON Netherlands Institute for Space Research, Niels Bohrweg 4, 2333 CA Leiden, The Netherlands}
 \affiliation{Optics Research Group, Department of Imaging Physics, Delft University of Technology, Van der Waalsweg 8, 2628 CH, Delft, The Netherlands}

\date{\today}

\begin{abstract}
Transition edge sensor (TES) microcalorimeters and bolometers are used for a variety of applications. \Mark{The sensors are based on the steep temperature-dependent resistance of the normal-to-superconducting transition, and are thus intrinsically sensitive to magnetic fields.} Conventionally the detectors are shielded from stray magnetic fields using external magnetic shields. However, in particular for \Mark{applications} with strict limits on the available space and mass of an instrument, external magnetic shields might not be enough to obtain the required shielding factors or field homogeneity. Additionally, these shields are only effective for magnetic fields generated external to the TES array, and are ineffective to mitigate the impact of internally generated magnetic fields. Here we present an alternative shielding method based on a superconducting groundplane deposited directly on the backside of the silicon nitride membrane on which the TESs are located. We demonstrate that this local shielding for external magnetic fields has a shielding factor of at the least $\sim$ 75, and is also effective at reducing internal self-induced magnetic fields, as demonstrated by measurements and simulation of the eddy current losses in our AC biased detectors. Measurements of 5.9~keV X-ray photons show that our shielded detectors have a high resilience to external magnetic fields, showing no degradation of the energy resolution or shifts of the energy scale calibration for fields of several microTesla, values higher than expected in typical real-world applications.
\end{abstract}

\maketitle

\section{Introduction} \label{sec:intro}

Transition Edge Sensors (TESs) are among the most sensitive micro-calorimeters currently available. A TES is a cryogenic detector that can be used to measure the energy of a photon or particle with very high resolving power by utilizing the steep temperature-dependent resistance of the \Mark{superconducting-to-normal} transition \cite{Irwin1996}. They are often used as X-ray spectrometers, both for ground-based experiments such as HOLMES \cite{Puiu2018} and several setups at beamline facilities \cite{Doriese2017}, as well as for future space-borne systems such as X-IFU \cite{Barret2020} and HUBS \cite{Cui2020}. TESs for this application typically consist of a superconducting bilayer with a finely tuned critical temperature ($T_c$) and normal resistance ($R_n$) fabricated on a membrane to ensure sufficient thermal isolation between the bilayer and its surroundings. \Mark{A schematic overview of a TES is visible in Fig. \ref{figure:Array}(a).} In most applications, large numbers of TESs have to be \Mark{read out} simultaneously at cryogenic temperatures, requiring the \Mark{use} of multiplexing techniques. These can be either based on DC-biased readout of the detector such as Time Division Multiplexing (TDM) \cite{Doriese2016}, Code Division Multiplexing (CDM) \cite{Morgan2016}, and microwave SQUID multiplexing \cite{Nakashima2020}, or based on AC-biased readout such as Frequency Domain Multiplexing (FDM) \cite{Akamatsu2021}. Each readout scheme sets specific requirements on the highly tuned detectors in term of response times, resistance, and uniformity across the TES array.

Considering the challenging environments in which TES arrays are often operating, one of the most pressing issues to solve is the sensitivity of the TES to magnetic fields. The main reason for this sensitivity is that, because of the coupling between the TES bilayer and the higher $T_c$ leads of the electrical circuit, the TES acts like a weak-link, similar to a Josephson junction \cite{Sadleir2010, Kozorezov2011}. The application of magnetic fields on such a structure induces Fraunhofer-like oscillations in the (critical) current \cite{Smith2013, Gottardi2018}. Additionally, the magnetic field influences the steepness of the superconducting transition, typically parameterized using the dimensionless $\alpha = \frac{T}{R}\frac{\partial R}{\partial T}$ and $\beta = \frac{I}{R}\frac{\partial R}{\partial I}$, which in turn is related to important device properties such as the time-constants and energy resolution. Changes in the pixel responsivity can lead to shifts of the energy scale calibration \footnote{\label{Vaccaro}D. Vaccaro et al., submitted to Journal of Low Temperature Physics}. The TESs are particularly sensitive to magnetic fields perpendicular to the plane of the bilayer. The sensitivity to parallel magnetic fields is much smaller primarily due to the reduced cross-section in this direction \cite{Hijmering2013}. For perpendicular fields, only a few $\upmu$T is enough to have significant impact on the individual device performance \cite{Zhang2017}. When looking at the full-array level, spatial variations of the magnetic field could be even more detrimental. These spatial variations cannot be compensated by a field coil near the TES array, and they reduce the uniformity of the optimal operating parameters of the different pixels. This is a problem in particular for the DC-biased readout schemes in which multiple pixels are biased in series, which does not allow the individual tuning of pixels to their optimal operating point.

In a typical experimental setup or instrument there are many potential sources of stray magnetic fields that might cause issues for the performance of the TESs. In general we can divide them into two categories. The first is external magnetic fields originating from outside of the TES array and readout, \Mark{such as Earth's magnetic field}, stray fields from nearby equipment, or fields generated by mechanical cryocoolers or adiabatic demagnetization refrigerators \cite{Jackson2018}. In the particular case of X-IFU another source is the bias currents in the anti-coincidence detector located a very short distance below the TES array \cite{DAndrea2020}. The second type of fields is internal magnetic fields that arise from the electrical currents within the devices themselves, such as the current used to bias the TESs within the superconducting transition.

The magnitude of external magnetic fields is generally reduced using a combination of mu-metal and superconducting shields \cite{Bergen2016, Vavagiakis2021}. However, the use of these magnetic shields can be problematic, especially in space applications where there are strict limitations on the size and weight of an instrument. Additionally, the effectiveness of the external magnetic shields is always limited by the openings necessary to let in the light to be detected or for wiring or thermalization. This not only reduces the attainable shielding factor, but can also lead to spatial variations in the residual field. Considering the internal magnetic fields; these are inherently difficult to mitigate, as they are generated within or very close to the detectors. One proposed solution is the fabrication of devices with turn-around style electrical leads, where the return current follows the same path as the incoming current \cite{Ishisaki2008,Swetz2012}. However, this solution requires significant changes to the fabrication process, and has for this reason not been widely adopted in high performance TES arrays.

In this work, we have followed an alternative approach to mitigate the effects of both internal and external magnetic fields, namely by using an on-chip superconducting layer to reduce the magnitude of magnetic fields perpendicular to the TES bilayer. The application of superconducting layers for this purpose has been done before, embedding the layer in the membrane or making it part of the wiring \cite{Luukanen2003, Ishisaki2008} or burying it in the substrate \cite{Finkbeiner2011}. An alternative approach was taken by Harwin \textit{et al.} \cite{Harwin2020}, who fabricated an array of micro-pillars capped with a superconducting film, designed such that the pillars would fit within the silicon grid of the membranes of the TES array. Here, we extend on these previous studies, and shield our detectors using a superconducting groundplane (SGP) located underneath the TES at the backside of the silicon nitride membrane. The advantage of this method is that the deposition of this layer can be done on fully finished arrays, requiring no changes to the fabrication process. At the same time, it allows us to place the SGP very close to the TESs, improving the shielding factor, with the spacing given by the thickness of the silicon-nitride membrane. We have directly compared devices with and without SGP on the same chip. We have observed no impact on important detector properties, implying that the SGP can be applied to TES arrays without requiring alterations to the readout circuit. At the same time, we demonstrate the effectiveness of the shielding to both external magnetic fields as well as to self-generated internal magnetic fields.

\section{TES Array} \label{sec:setup}

\begin{figure*}
\centering
\includegraphics[width=\linewidth]{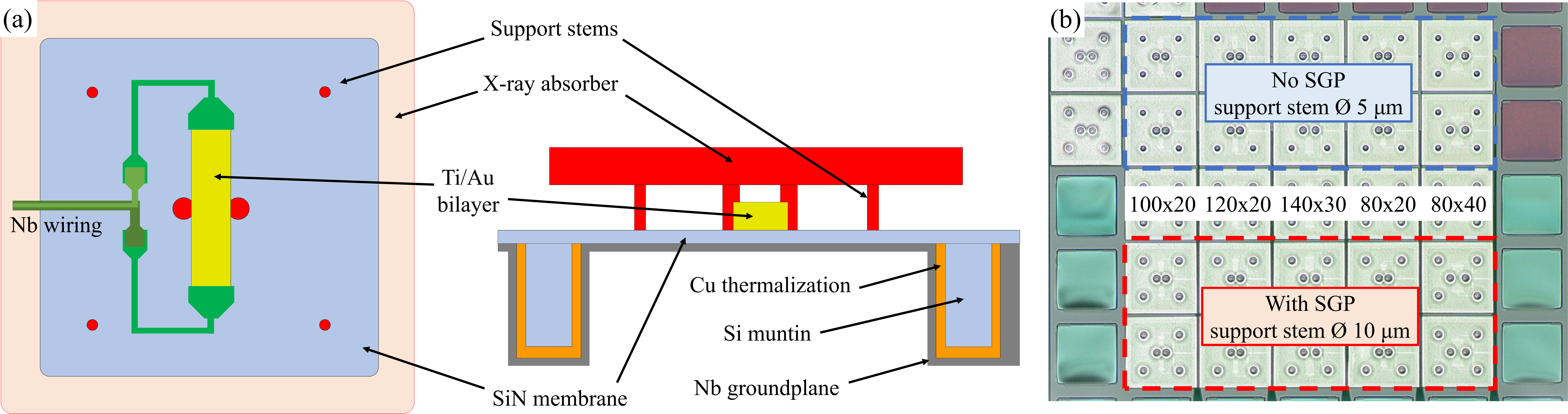}
\caption{(a) Schematic of the top-view (left) and side-view (right) of the TES detectors, showing the bilayer (yellow), niobium wiring (green), X-ray absorber (red), silicon-nitride membrane and silicon grid (light-blue), copper thermalization layer (orange), and SGP (grey). (b) Microscope image of the 5x5 TES array. Each column contains five devices of a different geometry as indicated by the labels in the center row (length $\times$ width in units micrometers). The SGP is present on the backside of the bottom three rows (visible as the blue-greenish color at the edges of the array). The top two rows do not have the SGP and serve as a reference.}
\label{figure:Array}
\end{figure*}

We have studied a selection of pixels on a $5 \times 5$ array containing five different TES designs of varying geometries and aspect ratios. The TESs consist of a rectangular Ti/Au bilayer with a sheet resistance of about 26~m$\upOmega$/$\square$ and $T_c$ between 93 and 96~mK, depending on the geometry. The TESs are connected to niobium leads and coupled to 240$\times$240~$\upmu$m$^2$ gold X-ray absorbers. The absorbers have a thickness of 2.35~$\upmu$m and are placed at a height of 3.5~$\upmu$m above the TES. The TESs are fabricated on top of a 0.5~$\upmu$m thick silicon-nitride membrane for thermal isolation between the TES and the thermal bath. Typical values for the thermal conductance between the bilayer and the thermal bath are 50~-~120~pW/K, depending on the TES geometry. Extensive measurements of the fabrication, properties, and performance of this type of array are presented elsewhere \cite{Nagayoshi2019, Wit2020}. A schematic overview of the TES design is shown in Fig.~\ref{figure:Array}(a). There is a small variation in the design of the support stems between the top two rows of the array, which have a stem diameter of 5~$\upmu$m, and the bottom three rows, which have a diameter of 10~$\upmu$m. This change causes a reduction of the thermal conductance of the top two rows compared to the bottom three rows of about 20 pW/K \cite{Taralli2021}. This difference does not influence the results presented in the next sections.

A Nb film was deposited through the silicon grid on the backside of membrane by magnetron sputter deposition, \Mark{coating both the membrane and muntins with a layer of Nb}, as indicated in Fig.~\ref{figure:Array}(a). During the deposition the silicon grid acts as a collimator, limiting the angles with which the sputtered Nb strikes the backside of the membrane. The reduction of deposition rate due to the collimator effect was compensated for. The final film thickness is 65~nm with a low $T_c$ of approximately 4~K due to the sub-optimal deposition conditions. This was measured on a representative monitor structure for the deposition of the Nb on the backside of the array. \Mark{The thickness on the vertical surfaces have not been measured, but is expected to be less than 65~nm due to the partial shadowing of the deposition. The thickness of 65~nm was selected as a compromise between having a layer thick enough to have a sufficiently high $T_C$, and thin enough to minimize the risk of introducing stress in the membrane or overheating the TES during deposition. Additionally, a thin layer keeps the parasitic thermal conductance and heat capacity as low as possible. A part of the array was covered during the deposition such that} the Nb film only covers the bottom three rows of pixels, leaving the top two rows without SGP to serve as a reference for comparison. This situation is clearly visible in Fig.~\ref{figure:Array}(b), in which the niobium film is visible at the edges of the array as the blue-greenish color at the backside of the silicon-nitride membranes.

For these proof-of-principle tests, the SGP was fabricated at the backside of a fully finished detector array. Based on comparison measurements between the devices with SGP and without SGP, the impact of adding the Nb on the most important properties of the TESs is negligible in the absence of magnetic fields. \Mark{Theoretical estimations indicate that the additional heat capacity is negligible and the parasitic thermal conductance between the TESs and the thermal bath is very small. Indeed, measurements of the critical temperature and thermal conductance show no difference between the pixels with and without SGP within the measurement uncertainty.} There also seems to be no significant impact on the shape of the superconducting transition ($\alpha$ and $\beta$), noise characteristics, or energy resolution. At low bath temperatures we measure an increased critical current for the pixels with SGP, presumably due to \Mark{an increased uniformity of the current \cite{Barone1982}.} However, for temperatures close to the $T_c$ of the bilayer the differences in the critical current between the pixels with and without SGP vanish. \Mark{Some exemplary data comparing devices with and without SGP can be found in the Supplemental Material \footnote{See Supplemental Material at [URL will be inserted by publisher] for theoretical estimations of the thermal impact of the SGP, as well as exemplary data comparing devices with and without SGP}.\label{SM}}

The array is mounted on a copper bracket at the mixing chamber of a dilution refrigerator. Magnetic shielding is achieved using a lead and cryoperm shield around the setup at the mixing chamber, and a mu-metal shield around the outside of the cryostat. These measures reduce the magnitude of stray magnetic fields at the TES array to $\ll 1 \upmu$T, as measured using the reference pixels. A small Helmholtz coil placed around the copper bracket can be used to apply a uniform magnetic field perpendicular to the TES array. All TESs are biased using an alternating current in an FDM readout system operated in single pixel mode. In this scheme, each pixel is connected to a superconducting LC resonator with a specific bias frequency between 1-5~MHz \cite{Bruijn2012}. This allows us to characterize many pixels in a single cryogenic run. Details of the FDM readout and measurement setup are given elsewhere \cite{Akamatsu2021, Wit2020}. In general the bias frequency used to read out the pixels influences the pixel performance due to the frequency-dependent weak-link effect. Therefore, when looking for effects of the SGP we compare pixels of the same geometry measured at similar bias frequencies (within $\sim$ 200~kHz). Furthermore, we will not study any devices in the center row of the array, since this row is located at the edge of the SGP where the quality of the coverage of the Nb layer is uncertain.

\section{Shielding External Magnetic Fields} \label{sec:External}

We start by investigating the effects of the SGP on external magnetic fields. For best operation of a TES, the component of the magnetic field perpendicular to the TES should be as close to zero as possible. Typically this is done using magnetic shielding (superconducting and high magnetic permeability materials) around the cryogenic setup and TES array in combination with a Helmholtz coil to tune the local magnetic field to zero. However, particularly for large TES arrays, the presence of local variations in the magnetic field cannot be fully prevented using these methods.

The effectiveness of the SGP to shield the TESs from external magnetic field can be expressed in terms of a shielding factor ($SF$), defined as the ratio between the magnitude of the residual magnetic field at the location of the TES with the SGP and the magnitude of the external magnetic field without SGP. Assuming a perfect Meissner state, this shielding factor is purely dependent on the geometry of the system. For simplicity, we ignore the complex corrugated shape of the SGP due to the silicon grid, and instead assume a flat superconducting disk with radius $a$. For that situation the axial component of the magnetic field at a height $z$ above the disk is given by Claycomb \textit{et al.} \cite{Claycomb1999}:
\begin{equation}
B(z) = B_0 \left[ 1 - \frac{2}{\pi} \left( \tan^{-1}{ \left( \frac{a}{z} \right)} - \frac{az}{a^2 + z^2} \right) \right] \approx B_0 \left[ \frac{4}{\pi}\frac{z}{a} \right],
\end{equation} 
where the right-hand side of the equation is valid in the limit $a \gg z$. For our system $z = 0.5 \upmu$m, as the separation between the TES and the SGP is given by the thickness of the membrane. \Mark{Defining an effective radius $a$ of the SGP is not so straight-forward for our array. A conservative estimate is made by assuming only the SGP below the membrane itself contributes to the screening, in which case $a/z \sim 200$. Thus from a purely geometric point of view, the residual field at the TES is attenuated by a few orders of magnitude. In practice, the attenuation must be limited by the London penetration depth, defects in the Nb film, and flux-focusing effects \cite{Brandt1998}.}

\begin{figure}
\centering
\includegraphics[width=\columnwidth]{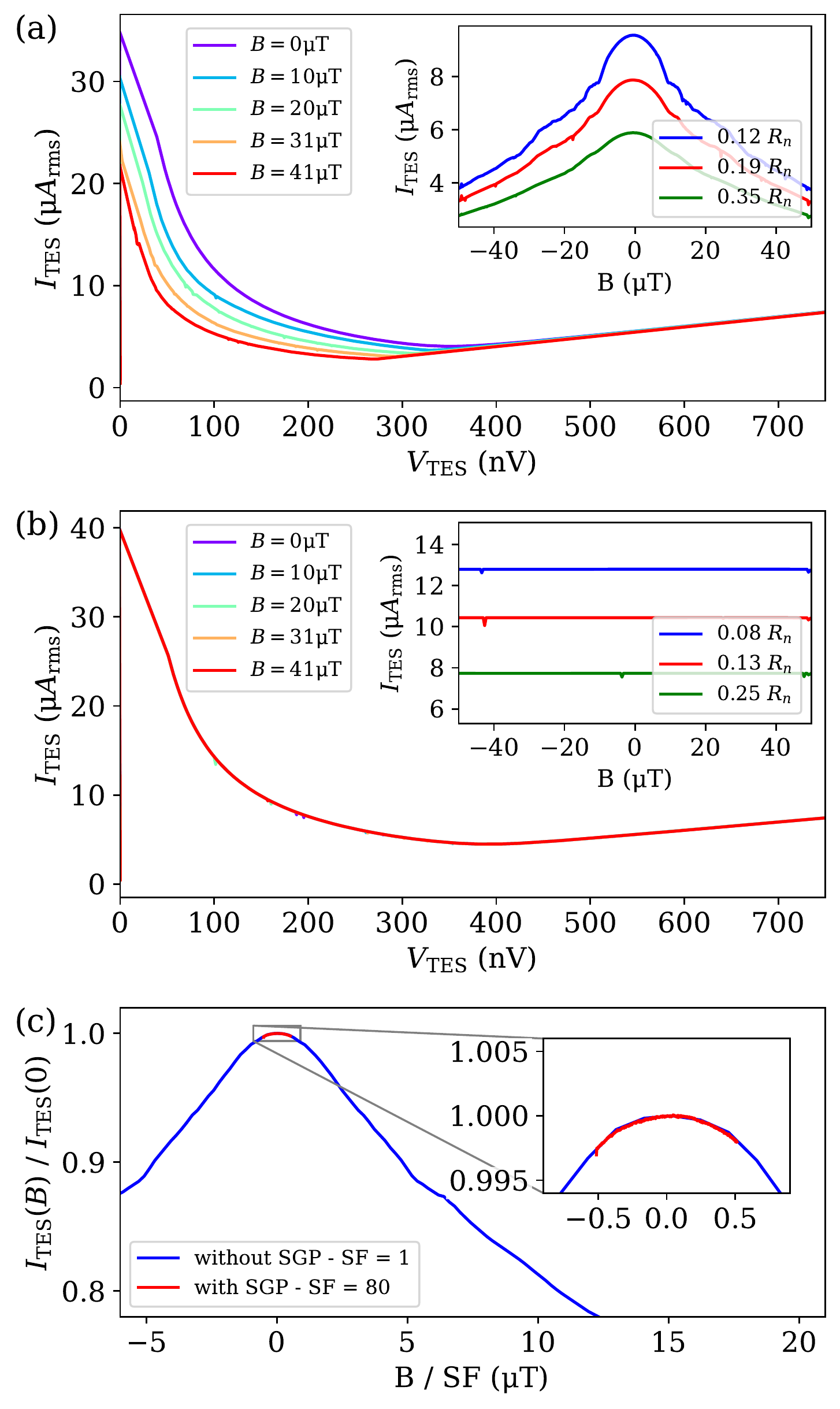}
\caption{Calibrated IV curves for applied magnetic fields ranging from 0~to~41~$\upmu$T, indicated by different colors, where (a) shows the data for a standard $80\times20~\upmu$m$^2$ pixel without SGP, and (b) for a pixel with SGP. The inset show the measured TES currents at three constant bias voltages (300, 400, and 500 mV) as a function of applied magnetic field. (c) TES current versus applied magnetic field divided by the shielding factor $SF$ for two $80\times40~\upmu$m$^2$ pixels, one without SGP (blue) and one with SGP (red).}
\label{figure:B-shielding}
\end{figure}

We characterize the shielding to external magnetic fields provided by the SGP using measurements of the TES current-voltage characteristic (IV curve) for a number of applied magnetic fields, shown in Fig. \ref{figure:B-shielding}. Fig. \ref{figure:B-shielding}(a) shows the IV curves for a pixel without SGP as a reference. The external magnetic field effectively lowers the $T_c$ of the bilayer, reducing the required power to bias the TES in the transition. As a result, the IV curve shifts downwards, as visible in the main figure. In the inset, the magnetic field dependence of the current in the TES ($I(B)$) is visible for three different bias points defined by the resistance at $B = 0~\upmu$T. A Fraunhofer-like pattern is visible resulting from the interaction between the magnetic field and the TES bilayer acting as a weak-link under influence of the niobium leads. Now let us compare these results with Fig. \ref{figure:B-shielding}(b), which shows the IV curves for one of the pixels with SGP. Whereas for the unshielded pixel the magnetic field clearly induces a shift of the IV curves towards lower power, no significant reduction in the detector power is observed for the shielded pixel (visible from the overlap of the IV curves measured at different fields). The measured $I(B)$ curve shown in the inset confirms the strongly reduced impact of the external magnetic field on the superconducting state of the TES.

To obtain an estimate for the shielding factor for external magnetic fields, we can look at the field dependence of the TES current of the most sensitive pixel design \Mark{($L 
\times W = 80 \times 40~\upmu$m$^2$)}. This data is shown in Fig. \ref{figure:B-shielding}(c). For this type of pixel, at the highest applied magnetic fields the pixel with SGP (red line) shows a small reduction in the TES current of $\sim 0.2$ \% at $B \sim 40~\upmu$T. By comparing this data to a similar pixel without SGP (blue line) we find which magnetic field is required to achieve the same relative reduction in the TES current. From this we can estimate the effective field felt by the TES, given by the $B/SF$. The best match between the pixels is found for $SF \sim 80$, indicating that the SGP can reduce the magnitude of external magnetic fields by nearly two orders of magnitude.

An important remark concerning the external magnetic fields and the use of an SGP is that it is very important to cool the setup in zero field. When the SGP transitions from the normal to the superconducting state in the presence of a magnetic field, this magnetic field can be trapped even after the external field is removed. Trapping of magnetic flux in SGP structures has been reported before \cite{Ishisaki2008, Finkbeiner2011}. We have tested the effect of field cooling by warming up the full setup to approximately 10K, a temperature well above the $T_c$ of the niobium film. At this temperature, we applied an external magnetic field using the Helmholtz coil which was maintained during the cooldown to a base temperature of $T = 50$~mK. After removal of the external magnetic field, the residual field due to flux trapping was determined by measuring $I(B)$ for the pixels without SGP. We observed a residual field of \Mark{$\sim$20-30~\%} of the field that was applied during the cooldown \Mark{(See the supplemental material for more information [30])}. Whether the only partial trapping of the applied field is due to the intrinsic properties of the film or a geometric effect is unknown. We assume the measured residual field for the pixels without SGP is representative for those with SGP. The trapped field can be removed by doing a thermal cycle above the $T_c$ of the SGP. In order to prevent trapping of flux in the film, zero-field cooling is required. The magnitude of the field that can be accepted during the cooldown depends on the sensitivity of the detectors and readout chain. For narrow devices (width 20$~\upmu$m or less) measured under AC bias, detectors with SGP can be cooled down in fields of up to 1-2$~\upmu$T without impact on the detector performance.

\section{Shielding Self-Induced Magnetic Fields} \label{sec:Internal}

The second category of magnetic fields that interact with the TESs is the self-induced magnetic fields. These fields are generated by the bias currents in the bilayer and leads. These self-induced fields can be particularly important when using low resistance devices for which relatively high bias currents are needed to bias the detectors within the superconducting transition. These self-fields are shown to reduce the steepness of the superconducting transition, reducing the detector sensitivity \cite{Swetz2012, Sakai2018}. Here we will show that the SGP is also an effective method to reduce the impact of self-induced fields.

The way in which the self-induced fields can be measured depends on the way in which the detectors are biased. In the case of DC biased detectors the presence of the self-induced field is visible as a skewing of the $I(B)$ curve proportional to the magnitude of the bias current. Related to this is the appearance of discontinuities in $\alpha$ and $\beta$, the magnitude of which depends on the geometric coupling between the bilayer and bias leads \cite{Smith2013}. Under AC biasing, it is more difficult to identify the effects of the self-induced fields. In general the TESs designed for operation under AC bias have a higher normal resistance \cite{Wit2020}, meaning typical operating currents (and hence self-induced fields) are smaller. Additionally the shift of the $I(B)$ curve can no longer be used as an indicator, since the self-induced field is alternated at MHz frequencies, smoothing out the effects. However, we are able to see the presence of self-induced fields via their interaction with normal metal in the vicinity of the leads and bilayer: the alternating fields induce eddy currents in these metal structures leading to AC losses \cite{Gottardi2017, Sakai2018}. These losses are mainly located in the X-ray absorber as this is located at merely $3.5 \upmu$m above the plane of the TES. The AC losses are measured as an additional parasitic resistance in the LC-resonator circuit. While for our devices these AC losses are sufficiently small that they do not affect the detector performance, here we will use them to demonstrate the effectiveness of the SGP to reduce the impact of self-fields. 

\begin{figure}
\centering
\includegraphics[width=\columnwidth]{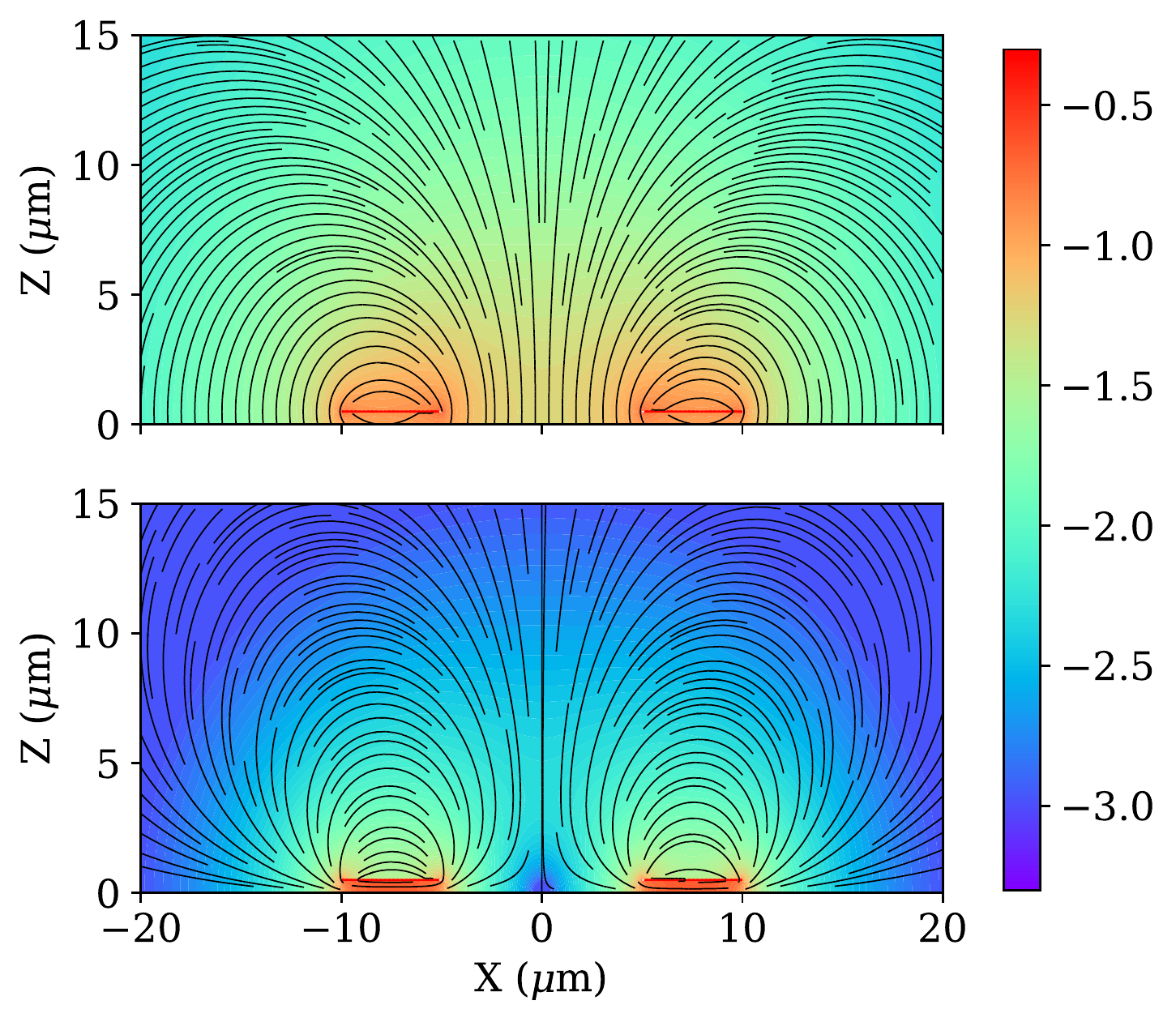}
\caption{Magnetic field distribution of two current carrying wires with (top) and without (bottom) SGP. In the latter case, the wires are placed 0.5~$\upmu$m above the SGP. The colors indicate the logarithmic magnitude of the magnetic field per unit current. Values near the wires are capped at $0.5$~T/A.}
\label{figure:B_calc}
\end{figure}

We can theoretically estimate the effect of the SGP on self-induced fields using the method of images (assuming perfect Meissner screening)\cite{Jackson2018}. We calculate the magnetic field from two current-carrying wires representing the bias leads, placed at a height $h$ above an infinite superconducting plane placed at $z = 0$. The magnetic field distribution above the SGP is then given by the superposition of the original current source and an image current flowing in the opposite direction at a height -$h$ underneath the SGP. The field produced by the screening current is identical to the field emanating from the image current \cite{Claycomb1999}. The field of each wire is calculated using the Biot-Savart law. In Fig. \ref{figure:B_calc} we illustrate the resulting field distribution with (top) and without (bottom) the SGP. The magnetic field without SGP resembles that of a magnetic dipole, while the SGP effectively changes this into a quadrupole field which falls off at a much faster rate as the distance to the wires increases.

The precise attenuation of the field due to the presence of the SGP depends on the separation between the SGP and the current carrying wire. For the simplest case in which we only consider the field at a height $z$ directly above a single wire, the attenuation $\Gamma$ is given by:
\begin{equation}
\Gamma = \frac{B_{w.SGP}}{B_{w/o.SGP}} = 1 - \frac{(z-h)}{(z+h)}
\end{equation}
Here $\Gamma = 1$ corresponds to zero attenuation, while $\Gamma = 0$ corresponds to total attenuation of the field. At the bottom of the absorber (corresponding to $z = 4~\upmu$m with $h = 0.5~\upmu$m), this leads to $\Gamma = 0.22$. A higher shielding factor can be attained by fabricating the SGP directly underneath the TES instead of on the backside of the silicon nitride membrane, or alternatively by increasing the height of the absorber.

\begin{figure}
\centering
\includegraphics[width=\columnwidth]{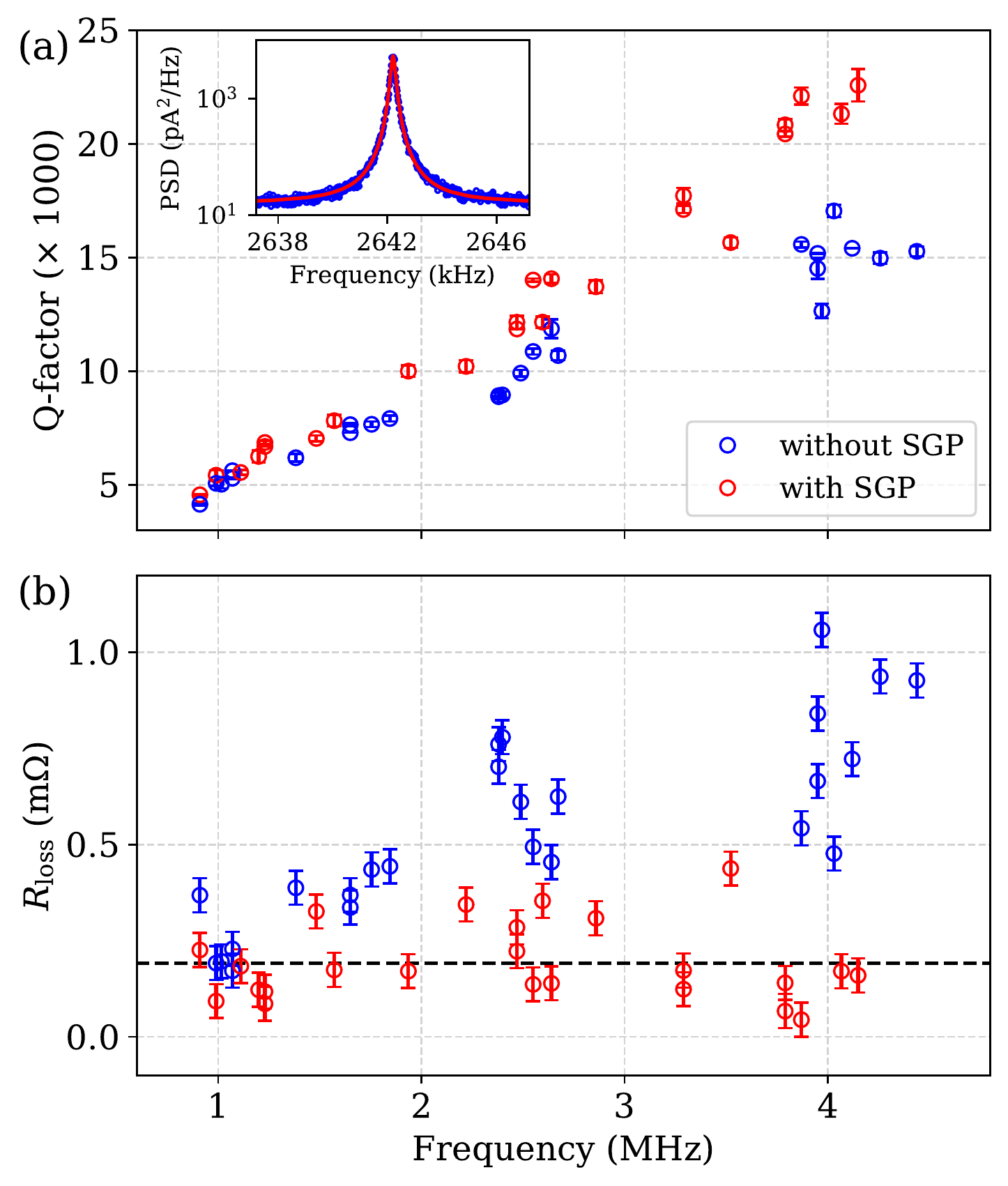}
\caption{(a) Q-factor obtained from the Lorentzian fit for each LC resonator, plotted as a function of the pixel bias frequency. (b) Extracted loss-factor for all pixels as a function of the bias frequency. Blue data indicates pixels without SGP, red data indicates pixels with SGP. The black dashed line is the estimated level of the residual losses in the electrical circuit. Several pixels where remeasured using different resonators to increase the number of available frequencies for each geometry.}
\label{figure:ACLosses}
\end{figure}

We experimentally estimate the AC losses by measuring the damping of each of the LC resonators coupled to a detector. \Mark{This is done by acquiring the SQUID power spectral density (PSD) with all the TESs in the superconducting state. The only excitation is the thermal noise in the LC resonator. In this way, the measured power spectrum reflects the transfer function of the LCR circuit formed by the LC resonator in series with any resistance. The contribution to the PSD of each resonator is fitted with a Lorentzian function to extract the Q-factor and resonance frequency $\omega_0 = 2\pi f_0$. The measured Q-factor for each peak is shown in Fig. \ref{figure:ACLosses}(a), with in the inset an example fit of a Lorentzian to the PSD for one of the resonators.} The \Mark{Q-factors for the different pixels range from} 5\,000 to 20\,000, increasing with the bias frequency due to the decreasing C. For the highest frequencies we see a saturation as the AC losses increase with frequency. The Q-factors of the pixels with SGP (red) are slightly higher than those of the pixels without SGP (blue). The total damping resistance in the RLC circuit is obtained from the measured parameters using $R_{tot} = \omega_0 L / Q$ (with $L = 1.17~\upmu$H the effective inductance). In Fig. \ref{figure:ACLosses}(b), we show $R_{loss}$, the residual losses in the LC-TES circuit after subtracting the impact of the shunt resistor, given by:
\begin{equation}
R_{loss} =  R_{tot} - R_{sh} \left(\frac{C}{C_{bias}} \right)^2
\end{equation}
Here $R_{sh} = 0.72 \Omega$, and $C/C_{bias} = 25$. The obtained $R_{loss}$ is the sum of the AC losses in the TES due to the self-fields and the residual losses in the electrical circuit, such as for instance the LC resonators. The fact that these two loss factors are both intrinsic to the LC-TES circuit means they are difficult to separate. However, we distinguish between the two using their different dependencies on the bias frequency; while the losses in the electrical circuit are expected to be frequency independent (at least for this limited frequency range), the AC losses increase with the square of the bias frequency.

We estimate the residual losses in the electrical circuit to be $0.19 \pm 0.10$ m$\Omega$ based on the $R_{loss}$ measured for the pixels with SGP. This level of losses is indicated in Fig. \ref{figure:ACLosses}(b) by the black dashed line. The remaining losses are attributed to the eddy current in the normal metal absorber. For the pixels with SGP, we see increasing losses as the frequency increases, while this is not observed in the pixels with SGP, which show a roughly constant value for all pixels.

\begin{figure}
\centering
\includegraphics[width=\columnwidth]{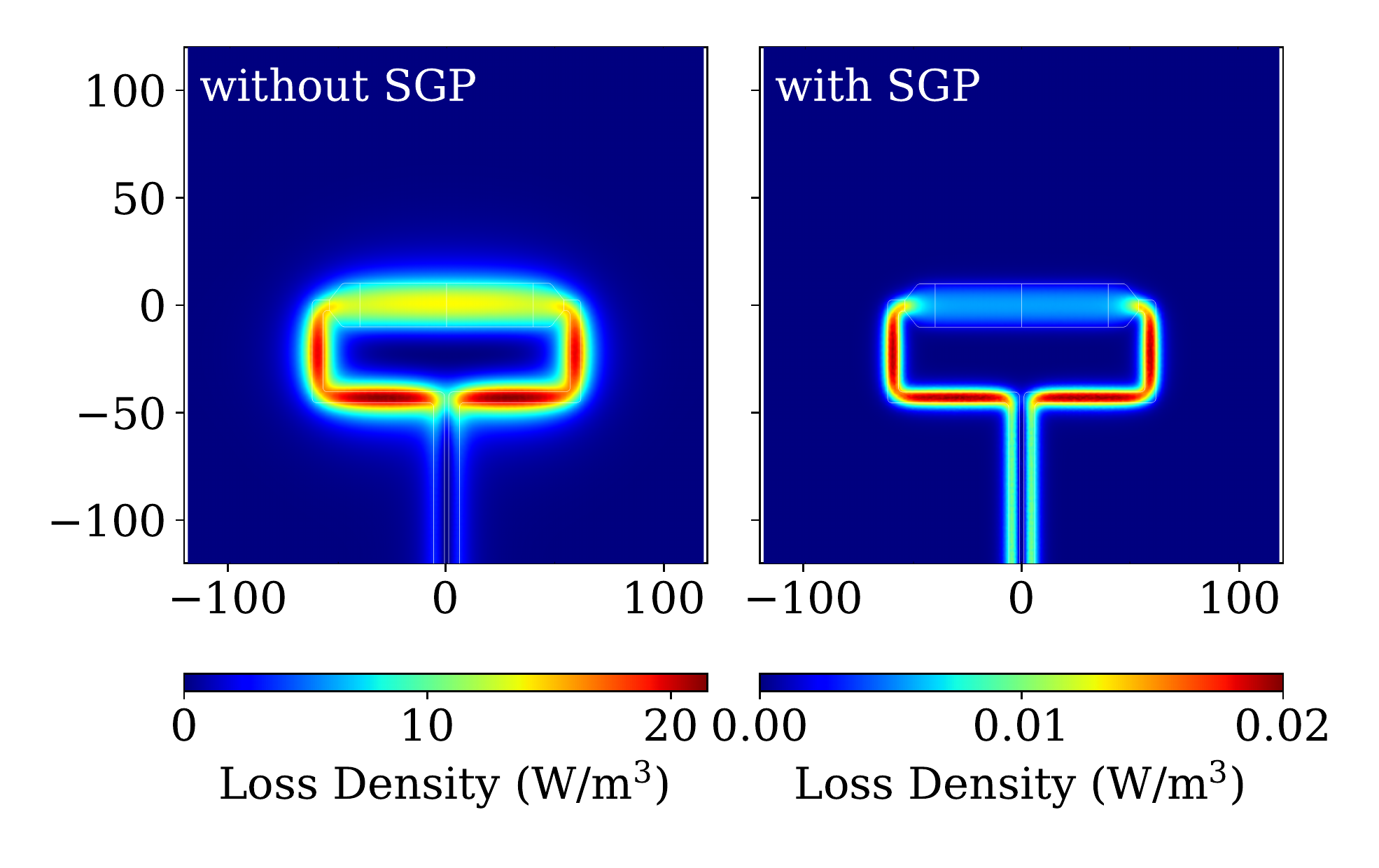}
\caption{FEM model heatmaps of the volumetric loss density at the bottom of the X-ray absorber due to the bias current in the leads and TES bilayer. Integration over the full volume of the absorber gives the total loss power $P_{loss}$. The left image shows the loss density for a detector without SGP, whilst the loss density for a detector with SGP is shown on the right. Note the different range of the colorbar in both figures.}
\label{figure:Heatmaps}
\end{figure}

We have used Finite Element Method (FEM) modeling to simulate the expected losses as a function of the bias frequency, both with and without the SGP. In Fig. \ref{figure:Heatmaps} we show an example of the calculated volumetric loss density in the absorber for a $80 \times 20~\upmu$m$^2$ TES measured at a bias frequency of 2 MHz. The geometry of the TES, wiring, and absorber all match those of the actual devices as outlined in Sec. \ref{sec:setup}. The only parameter that is varied in the simulation is the electrical conductivity $\sigma$ of the gold of the absorber, which for this figure was set to $0.5 \cdot 10^9$ S/m, a reasonable value for our gold films at low temperatures \Mark{\cite{Sakai2018, Wakeham2019}}. Integration over the full volume of the absorber gives the total loss power $P_{loss}$, from which the loss factor $R_{loss}$ can be calculated by dividing by the squared bias current. Note that in practice $R_{loss}$ \Mark{does not depend on} the bias current and only depends on the bias frequency, absorber conductance, and the geometry of the TES and wiring. The SGP is simulated as a plane located $0.5 \upmu$m below the TES with boundary conditions such that the normal component of the magnetic field must be zero at the interface. The field distribution arising from the FEM simulation closely matches the analytical results obtained from the method of images.

\begin{figure}
\centering
\includegraphics[width=\columnwidth]{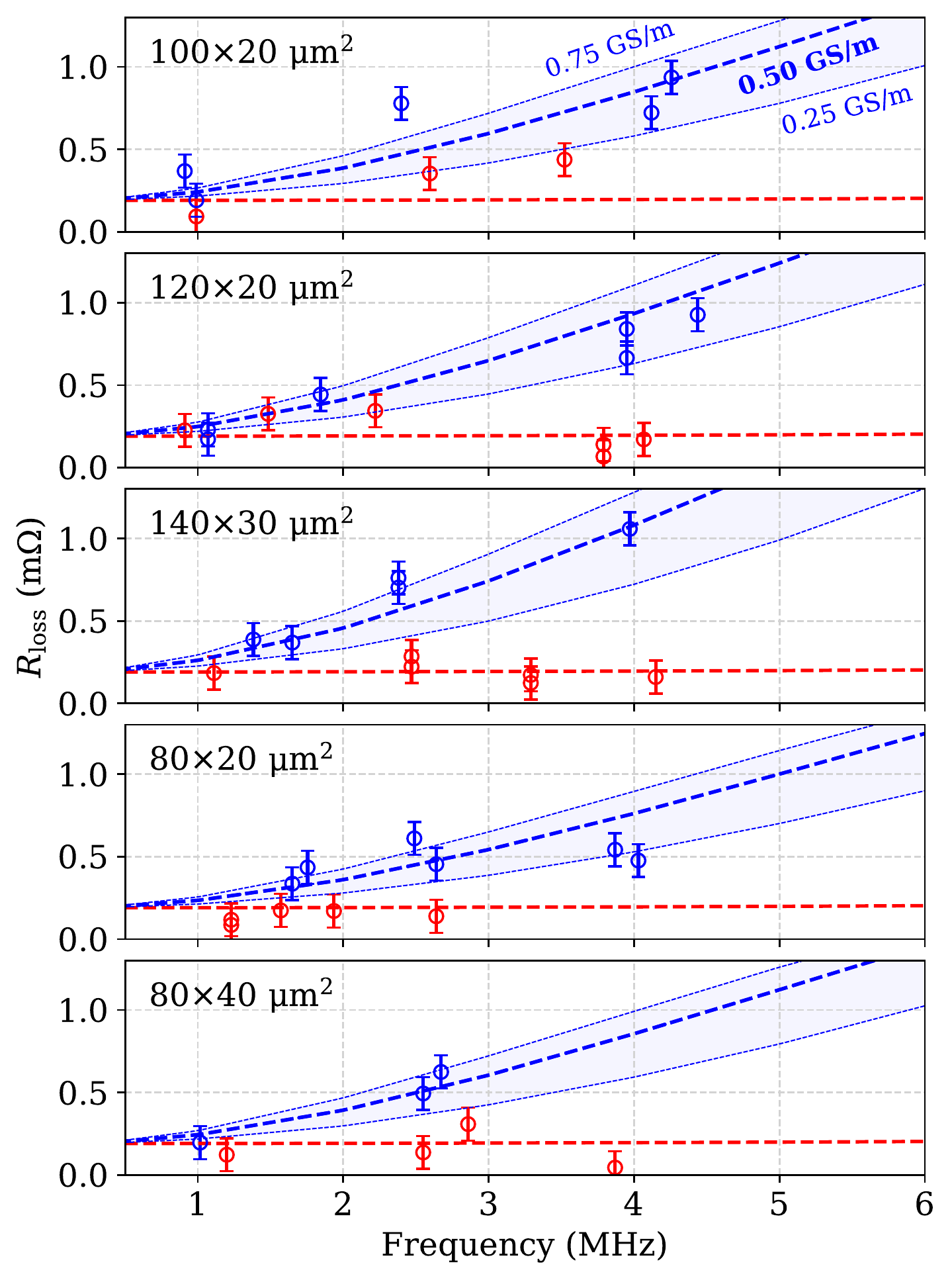}
\caption{$R_{loss}$ separated for the five different geometries. Blue data shows pixels without SGP, red data is pixels with SGP. The central dashed line is the result of FEM modeling assuming $\sigma = 0.50$ GS/m, while the blue area marks the region between $\sigma = 0.25$ GS/m and $\sigma = 0.75$ GS/m. The errorbars are given by the scatter in the losses measured for the pixels with SGP, which is believed to be dominated by the parasitic losses in the electrical circuit.}
\label{figure:Rloss_Sim}
\end{figure}

Fig. \ref{figure:Rloss_Sim} shows both the measured and simulated $R_{loss}$ for all pixels, separated to the different geometries. The central blue dashed line indicates the simulated losses assuming $\sigma = 0.5\cdot 10^9$ S/m, while the filled area around it marks the range between $\sigma = 0.25\cdot 10^9$ - $0.75\cdot 10^9$ S/m. The measured data for the pixels without SGP seems reasonably well explained by the losses in the absorber, even though the scatter in the data is too large to determine the $\sigma$ of the absorber with high accuracy. \Mark{For the pixels with SGP}, the total losses are dominated by the residual losses in the electrical circuit, as the FEM results confirm that the AC losses in the absorber never exceed $\sim 10~\upmu \Omega$. Thus, we again find that the SGP has the potential to significantly reduce the impact of self-induced magnetic fields on the TESs.

Note that we observe a clear deviation from the expected squared dependency of the losses on frequency. This is due to the skin effect; for the higher bias frequencies and conductance of the absorber, the skin depth $\delta = \sqrt{2 / \omega \sigma \mu}$ (with $\mu$ the magnetic permeability of gold) becomes similar or smaller than the thickness of the absorber, meaning the absorber effectively shields itself from the self-induced magnetic fields, reducing the integrated losses. This effect is also visible in the data presented by Sakai \textit{et al.} \cite{Sakai2018}.

\section{Gain and Energy Resolution}

The ultimate figure of merit for the TES is the energy resolution. Magnetic fields are known to affect the main detector properties, which could reduce the intrinsic sensitivity of the detectors. The fields also affect the gain of the detectors which results in a shift of the energy scale calibration. We assess the energy resolution of our devices by exposing the TES array to 5.9 keV X-rays originating from a $^{55}$Fe calibration source with a typical count rate of approximately 1 count per second per pixel.

\begin{figure}
\centering
\includegraphics[width=\columnwidth]{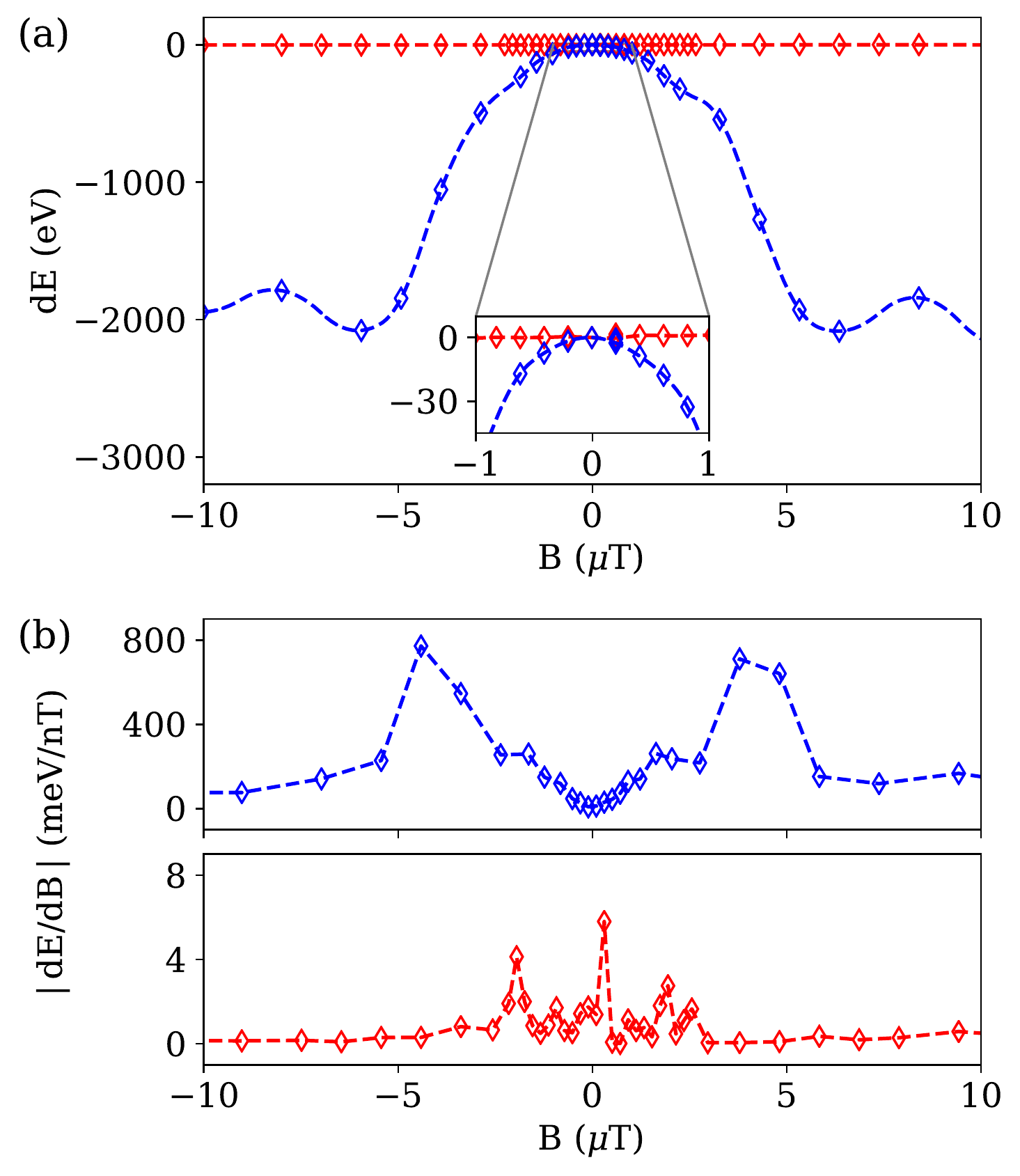}
\caption{(a) Shift of the measured energy of 5.9~keV X-rays for two pixels ($140 \times 30~\upmu$m$^2$), one without (blue curve) and one with (red curve) SGP. The inset shows a zoom in of the same data, focused on the region near $B = 0$. The dashed lines are guides to the eye. The filter template is calculated from the X-rays and noise measured at $B = 0$. (b) Absolute value of the local derivative of the data shown in (a).}
\label{figure:Gain}
\end{figure}

To characterize the sensitivity of the gain to changes in the magnetic field environment, we follow the method as outlined by Vaccaro \textit{et al.} [14]: We acquire about 500 X-ray and 500 noise events for a range of applied magnetic fields. The energy of the collected X-rays is determined using the optimal filtering technique, which aims to maximize the signal-to-noise ratio by appropriately weighing the various frequency components of the pulses \cite{Szymkowiak1993}. For every magnetic field value we analyze the pulses using a single filter template calculated for the events collected at $B = 0$. We then determine the shift of the energy $dE = E - E_0$, with $E$ the calculated pulse energy from the pulse and filter template, and $E_0$ the known position of the K$\alpha_1$ line at 5898.75~eV. In this way, for each magnetic field we can determine the shift of the K$\alpha_1$ line caused by the field-induced variations in the TES gain.

Fig. \ref{figure:Gain}(a) shows the measured shift of the energy for two pixels, one without (blue) and one with (red) SGP. For the pixel without SGP a big shift in the calibrated energy is observed as the magnetic field is increased. The inset shows a zoom in of the data close to $B = 0$. From the local derivative as shown in Fig. \ref{figure:Gain}(b) we find around $B = 0$ a gain scale sensitivity at a level of several tens of meV/nT, and a maximum sensitivity of $\sim$ 0.8~eV/nT near $B = \pm~4~\upmu$T. In contrast, looking at the data obtained from the pixel with SGP, the remaining magnetic field sensitivity is on the order of few meV/nT over the full magnetic field range, meaning no shift is observed within the statistical error of the energy measurement with only 500 counts. \Mark{Similar data for the other available geometries can be found in the supplemental material [30].}

\begin{figure}
\centering
\includegraphics[width=\columnwidth]{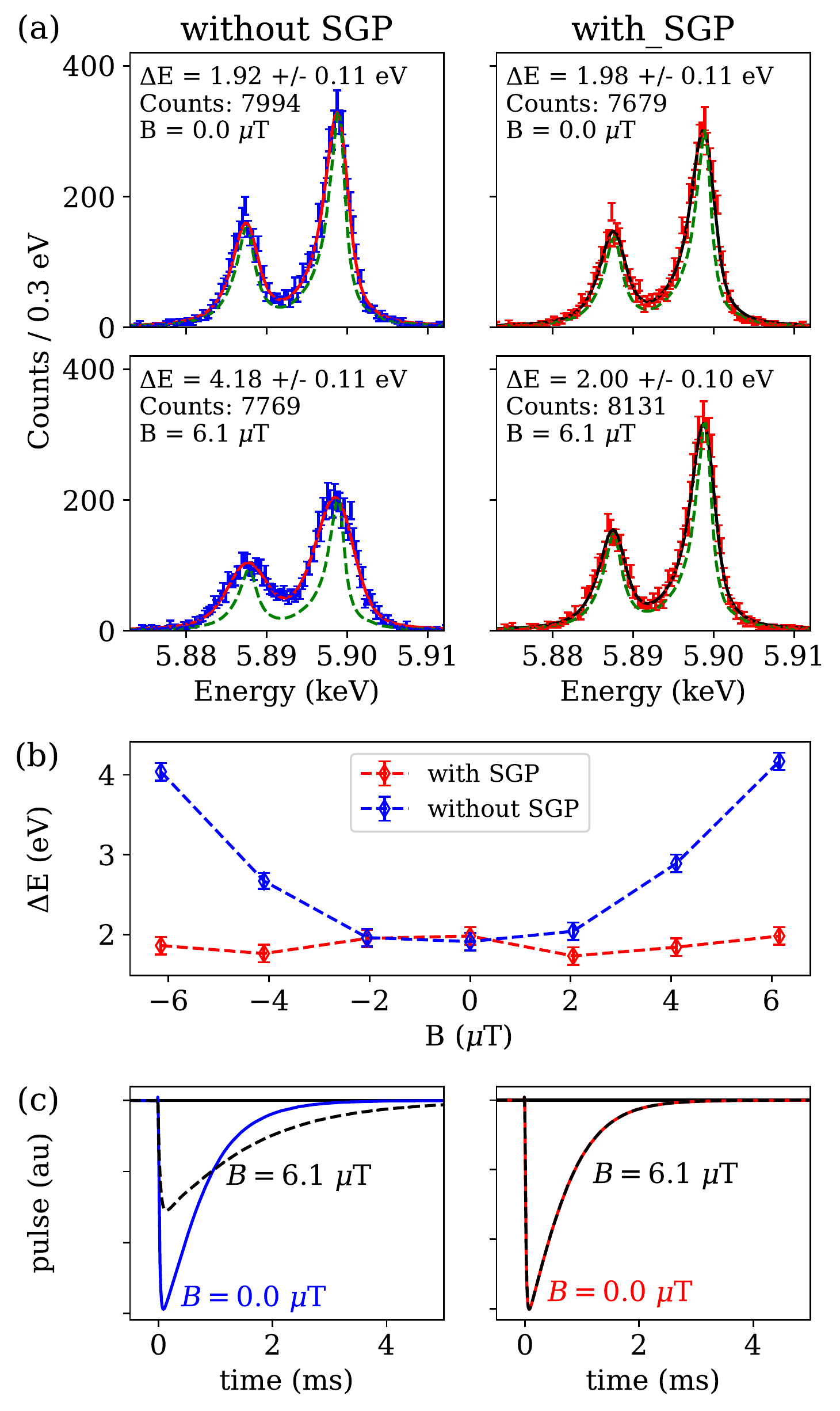}
\caption{(a) Example X-ray spectra of the Mn K$\alpha$ lines for $140 \times 30~\upmu$m$^2$ pixels without SGP (left column, blue data) and with SGP (right column, red data) at $B = 0$ (top row) and $B = 6.1 \upmu$T (bottom row). (b) Energy resolution at 5.9 keV versus applied magnetic field for two pixels, one without (blue curve) and one with (red curve) SGP. The energy in both (a) and (b) is calculated using the optimal filter template for each field value. (c) Example pulse shapes at $B = 0$ and $B = 6.1~\upmu$T for the pixels without (left) and with SGP (right).}
\label{figure:Energy_Res}
\end{figure}

To demonstrate the impact of magnetic fields on the intrinsic detector performance, and how the SGP also mitigates this effect, we have measured a high number of X-ray events (approximately 10000 per pixel per setting) for a number of applied magnetic fields. At each magnetic field value, the data is analyzed using the optimal filter template based on noise events measured at that specific field (as opposed to always using the template calculated at $B = 0$ such as was done for Fig. \ref{figure:Gain}). In this way the energy resolution is not affected by changes in the calibration of the energy scale, but is only determined by the pure detector properties. Example X-ray spectra of the Mn K$\alpha$ lines are shown in figure \ref{figure:Energy_Res}(a) for $B = 0$ and $B = 6.1~\upmu$T. The extracted intrinsic energy resolutions of the pixels as a function the applied field are shown in Fig. \ref{figure:Energy_Res}(b).

Both pixels achieve similar energy resolutions around 2~eV for the selected bias points. As the applied magnetic field is increased, the energy resolution degrades for the pixel without SGP (blue) quickly degrades to over 4~eV at $B = \pm~6~\upmu$T, while the shielded pixel (red) achieves the same energy resolution within the statistical error ($\pm~0.11$~eV). The degradation of the energy resolution for the unshielded pixel is accompanied by an increase in the pulse fall-time from 0.6~ms at $B = 0$ to 1.5~ms at $B = \pm~6~\upmu$T. See Fig. 
\ref{figure:Energy_Res}(c) for example pulse shapes at the different fields. The increasing fall-time indicates a smaller loop-gain of the electro-thermal feedback due to a reduction of both the TES power and $\alpha$. Fig \ref{figure:Energy_Res} clearly shows that even for these narrow devices measured under AC bias, which are relatively insensitive to magnetic fields when compared to the low-resistance, broad devices typically used for DC biased readout, small magnetic fields of only few microTesla are enough to significantly affect the TES performance. However, the SGP highly reduces the susceptibility of the pixels to magnetic fields, relieving the strict requirements set for external magnetic shielding.

\section{Conclusions}

To summarize, we have studied in what capacity the presence of a superconducting groundplane affects the magnetic field sensitivity of TES X-ray calorimeters. We have shown that the SGP provides mitigation to the effects of both external and self-induced magnetic fields, allowing the TES to retain their optimal performance even in the presence of magnetic fields \Mark{of} several tens of microTesla. This observation is of great importance to practical applications of TES arrays in situations with strong, non-uniform stray magnetic fields and strict limitations on the available mass or volume for magnetic shielding, such as in space applications. The SGP has the potential to give additional shielding to magnetic fields with a shielding factor of up to two orders of magnitude without adding fabrication complexity. \Mark{Future experiments have to confirm whether the performance of the SGP scales well when increasing the size of the TES array}.

One important point to stress is the necessity to cool the SGP in zero-field, as discussed in Sec. \ref{sec:External}. The possibility of flux trapping in these structures has been reported before, and was confirmed experimentally for this TES array. In the case of trapping, a thermal cycle above the critical temperature of the SGP is required to remove the fields. This means that for most applications the SGP alone will not be enough to be able to properly operate the TESs. Near zero-field cooling has to be achieved by combining the SGP with (a lightened version of) traditional low-temperature magnetic shielding based on mu-metal and superconducting materials external to the TES array \cite{Bergen2016}.

The shielding capability of the SGP increases when the distance between the TES and the SGP is decreased, which means that in principle integration of the SGP in the array fabrication is preferable for optimal shielding. However, here we show that the SGP can even be effective when deposited on the backside of an existing array. In this case, it provided significant shielding to magnetic fields without altering any of the vital detector properties in a significant way. This means that the SGP can be used as an easy to implement, cost-effective method to mitigate the magnetic field sensitivity of TES arrays without requiring a careful re-tuning of the full readout circuit.

\section{Acknowledgement}
The authors thank R. den Hartog for proofreading the manuscript. \Mark{SRON Netherlands Institute for Space Research is supported financially by NWO, the Dutch Research Council. This work was funded partly by NWO under the research programme Athena with Project No. 184.034.002 and partly by the European Space Agency (ESA) under ESA CTP Contract No. 4000130346/20/NL/BW/os.} It has also received funding from the European Union’s Horizon 2020 Program under the AHEAD (Activities for the High-Energy Astrophysics Domain) project with Grant Agreement Number 654215.

\section*{Data Availability}
The data that support the findings of this study are available from the corresponding author upon reasonable request.

\section*{References}

\bibliography{SG_bibliography}

\end{document}